\documentclass{ntmanuscript}
\title{Non-judgemental Dynamic Fuel Cycle Benchmarking}

\author{Anthony Micahel Scopatz$^1$}

\institute{$^1$University of South Carolina, Department of Mechanical 
    Engineering, Nuclear Engineering Program, Columbia, SC 29201}

\submitter{Anthony M. Scopatz}
\submitteraddress{541 Main Street, Columbia, SC 29208}
\submitteremail{scopatz@cec.sc.edu}


\usepackage{color}
\usepackage{graphicx}
\usepackage{booktabs} 
\usepackage{microtype} 
\usepackage{xspace}
\usepackage{listings}
\usepackage{textcomp}
\usepackage{ulem}
\usepackage{amssymb}
\DeclareMathAlphabet{\mathpzc}{OT1}{pzc}{m}{it}

\definecolor{listinggray}{gray}{0.9}
\definecolor{lbcolor}{rgb}{0.9,0.9,0.9}
\newcommand{\E}{\mathbb{E}}
\newcommand{\GP}{\mathpzc{GP}}
\newcommand{\LWR}{\mathrm{LWR}}
\newcommand{\FR}{\mathrm{FR}}
\newcommand{\Total}{\mathrm{Total}}
\newcommand{\argmin}{\mathrm{argmin}}
\newcommand{\CYCLUS}{\mathrm{Cyclus}}
\newcommand{\DYMOND}{\mathrm{DYMOND}}
\newcommand{\I}{\mathbf{I}}
\newcommand{\K}{\mathbf{K}}

\date{}
\begin{document}

\begin{abstract}
This paper presents a new fuel cycle benchmarking analysis methodology
by coupling Gaussian process regression, a popular technique in Machine 
Learning, to dynamic time warping, a mechanism widely used in speech 
recognition. Together they generate figures-of-merit that are applicable to
any time series metric that a benchmark may study. The figures-of-merit
account for uncertainty in the metric itself, utilize information
across the whole time domain, and do not require that the simulators
use a common time grid. Here, a distance measure is defined that can be used 
to compare the performance of each simulator for a given metric. Additionally, 
a contribution measure is derived from the distance measure that can be used 
to rank order the importance of fuel cycle metrics. Lastly, this paper 
warns against using standard signal processing techniques for error reduction.
This is because it is found that error reduction is better handled by the 
Gaussian process regression itself.

\end{abstract}

\section{Introduction}
\label{intro}
The act of fuel cycle benchmarking has long faced methodological issues 
as per what metrics to compare, how to compare them, and at what point in the
fuel cycle they should be compared 
\cite{wilson2011comparing,guerin2009benchmark,piet2011assessment}. 
The benchmarking mechanism described 
here couples Gaussian process models (GP) \cite{rasmussen2006gaussian} to 
dynamic time warping (DTW) \cite{muller}. Together these address how to 
generate figures-of-merit (FOM) for common nuclear fuel cycle benchmarking
tasks. 

Confusion in this area is partly because such activities 
are not in fact `benchmarking' in the strictest validation sense. Most fuel
cycle benchmarks are more correctly called code-to-code comparisons, or 
inter-code comparisons, as they compare simulator results. Importantly, 
there is an absence of true experimental data for a benchmark or simulator
to validate against. Moreover, the number of 
real world, industrial scale nuclear fuel cycles that have historically been 
deployed is not sufficient for statistical accuracy. Even the canonical 
Once Through fuel cycle scenario has only been deployed a handful of times.
For other advanced fuel cycles, industrial scale data is even more stark. 
Since fuel cycle simulation is thus effectively impossible to validate, 
non-judgmental methods of benchmarking must be considered. The 
results of any given simulator should be evaluated in reference to how 
it performs against other simulators in such a way that acknowledges that 
any and all simulators may demonstrate incorrect behavior. No simulator
by fiat produces the `true' or reference answer.

The other major conceptual issue with fuel cycle benchmarking is that there 
is no agreed upon mechanism for establishing a figure-of-merit for 
that is uniform across all fuel cycle metrics of interest. For example, 
repository heat load may be examined only at the end of the of the simulation,
separated plutonium may be used as a FOM wherever it peaks, and natural uranium 
mined might be of concern only after 100 years from the start of the 
simulation. Comparing at a specific point 
in time fails to take into account the behavior of that metric over time and 
can skew decision making. Additionally, the 
time of comparison varies based on the metric itself. This is a necessary 
side effect of picking a single point in time.
Furthermore, such FOMs are not useful for indicating why simulations differ, 
only that they do. Moreover, if such FOMs match, this does not indicate
that the simulators actually agree. Their behavior 
could be radically different at every other simulation time and converge
only where the metric is evaluated.  The one caveat 
here is that 
equilibrium and quasi-static fuel cycle simulators are sometimes able to 
ignore these issues because all points in time are treated equally.

Some dynamic FOMs do exist. However, these typically require that the metric
data be too well-behaved for generic comparison purposes. Consider the case 
of total power produced [GWe]. A FOM could be the sum over time of the relative error 
between the total power of a single simulator and the root-mean squared total power
of all the simulators together. However, such an FOM fails if the total power
time series have different lengths. Such differences could arise because 
of different time steps (1 month versus 1 year) or because of different 
simulation durations. Furthermore, suppose that a benchmark is posed as 
"until transition" in a transition scenario. It would defeat the purpose of 
the benchmark to force different simulators to have the same 
time-to-transition if they nominally would calculate distinct transition 
times. Therefore, a robust FOM should not impose the constraint of a uniform time grid.
 
The mechanisms used for benchmarking that have been discussed so far typically
do not incorporate modeling uncertainty coming from the simulator itself.
This is likely because most simulators do not compute uncertainty directly. 
Instead they rely on perturbation studies or stochastic wrappers around 
the simulator. Furthermore, metrics may add their own uncertainty from the 
data that they bring in (half-lives, cross-sections, etc.). 
However, even if such error bars were available for
every point in a time series metric, the traditional benchmark FOM 
calculations would ignore them.

The method described in this paper addresses all of the above issues. It 
creates FOMs that a dynamic fuel cycle benchmark will be able to use on any 
metrics of interest. Before going further, it is important to note that 
most fuel cycle 
metrics are time series and can be derived from the mass balances. 
Additionally, many metrics have an associated total metric that can be 
computed from the linear combination of all of its constituent features. 
For example, total mass flows are the sum of the mass flow of each nuclide
and total power generated is the sum of the power from each reactor type, 
such as light water reactors (LWR) and fast reactors (FR). These attributes 
are common to the overwhelming majority of fuel cycle metrics and so FOMs
may take advantage of such structures.

Gaussian process models are proposed here as a method to incorporate 
metric uncertainties and make the analysis non-judgmental with respect to 
any particular
simulator. Roughly speaking, a Gaussian process regression is a 
statistical technique
that models the relationship between independent and dependent parameters
by fitting the covariance to a nominal functional form, or kernel.
The kernel may have as many or as few fit parameters (also called 
\emph{hyperparameters}) as desired. One often used kernel is the squared 
exponential, or Gaussian distribution. Linear kernels and Gaussian
noise kernels are also frequently used alternatives. 

Using a Gaussian process is desirable because the model can be generated
from as many simulators as available and it will treat the results of
each simulators in the same manner. Unlike a relative error analysis, no 
simulator needs to be taken as the fiducial case. The Gaussian process model 
itself becomes the target to compare against. 
Moreover, the covariances do not need to be known to perform the benchmark.
They are estimated by the Gaussian process. Furthermore, once the 
hyperparameters are known, this can be used as a representative model for 
any desired time grid. Additionally, the incorporation of the uncertainties
in a Gaussian process are known to be more accurate (closer means) than 
assuming uncorrelated uncertainties.  The trade off is that the 
model is less precise (higher standard deviations) than the uncorrelated 
case \cite{hodlr}. Such a trade off is likely desirable because no simulator 
is necessarily more correct than any other simulator. For example, 
in an inter-code comparison, an outlier simulator may be the 
most correct, perhaps because it is higher fidelity than the others. Thus it
becomes important to quantify outliers rather than discard them.

However, a Gaussian process model of a metric for a set of simulators 
does not directly present itself as a FOM for that metric. Time series go 
into a Gaussian process and time series come out. The dynamic time warping 
technique is proposed as a method for deriving FOMs from the models.  

Dynamic time warping computes the minimal distance and path that it would 
take to convert one time series curve into another. This procedure is highly 
leveraged in audio processing systems, and especially in speech recognition
\cite{myers1980performance,muda2010voice}.
This is because the two time series that are being compared need not have
the same time basis.  It does not matter if one time series is longer than 
the other or if they have different sampling frequencies. The DTW distance
instead compares the shape the curves themselves. 

There is nothing about the dynamic time warping algorithm that is specific 
to speech recognition. The method
may be used in any time series analysis. For nuclear fuel cycle benchmarking,
the DTW distance can be used to compare the metric from each simulator to its
Gaussian process model. Using this as an FOM has the advantage of 
incorporating information from the whole time series, rather than just a 
specific point in the cycle.

Many benchmarking studies also wish to create a rank ordering of parameter
importance over all simulators. Examples of such benchmark questions include, 
``In a transition scenario,
which reactor is most important to the total generated power?'' and ``Which
nuclides are most important to the repository heat load?'' DTW distances 
of Gaussian process models of the constient parameters (e.g. the power from
each reactor type) to a Gaussian process model of the total (e.g. total 
generated power) can be used as a FOM itself or to derive other FOMs. 
This paper proposes a novel contribution metric. 
Contribution is taken to be a normalized version of the DTW 
distance for such rank ordering activities.

The remainder of this paper takes a narrative approach that walks through 
a ficticious example benchmarking activity. Generated power [GWe] data 
from DYMOND \cite{yacout2005modeling,feng2015dymond} and Cyclus 
\cite{DBLP:journals/corr/HuffGCFMOSSW15,cyclus_v1_0} is used throughout. 
The underlying fuel cycle being modeled is an
LWR to FR transition scenario that occurs over 200 years, starting from 2010.  
This data should be regarded as for demonstration purposes only. No deep
fuel cycle truths should be directly inferred and the data itself should be 
considered preliminary. However, 
using results from actual fuel cycle simulators shows how
non-judgmental benchmarking works in practice. Completely faked data could 
have been used instead, but this demonstration of the method is more 
believable.

In \S \ref{setup}, the problem is set up, mathematical notation is introduced,
and the raw data from the simulators is presented. In \S \ref{gp}, Gaussian 
process 
modeling is introduced. In \S \ref{dtw}, the dynamic time warping concept is
discussed. In \S \ref{contribution}, the novel contribution metric is 
derived.
\S \ref{filtering} warns against standard time series filtering
techniques. And finally, \S \ref{conclusion} contains concluding remarks
and ideas for future work in fuel cycle benchmarking and for 
using the mechanisms presented here.

\section{Benchmarking Setup}
\label{setup}

Suppose that there are $S$ simulators, indexed by $s$. The exact order 
of these simulators does not matter, however a consistent ordering should
be used. In the demonstration here, there are two simulators with $s=0$
being DYMOND and $s=1$ for Cyclus.

Furthermore, for any feature or metric there may be $I$ partitions, 
indexed by $i$. In this example benchmark, the total generated power metric
is studied and has constituent components of the power generated by LWRs 
($i=0$) and
the power generated by FRs ($i=1$).  Again, the ordering of these is not 
important, only that the ordering is consistent. An alternative example
that will not be examined here is the mass flow, which may be partitioned 
by its individual nuclides or by chemical element.

Now, denote a metric as a function of time $t$ for a given simulator and 
component as $m_s^i(t)$. For many metrics of interest 
to nuclear fuel cycle analysis, the following equality holds:
\begin{equation}
m_s(t) = \sum_i^I m_s^i(t)
\end{equation}
Thus $m_s(t)$ is the total metric over all constituent parts for a given 
simulator. This linear combination is useful when calculating contributions
(as seen in \S\ref{contribution}) but is not needed to compare various 
simulators to a Gaussian process model (see \S\ref{dtw}).

Additionally, call $u_s^i(t)$ the uncertainty of the metric 
$m_s^i(t)$. Note that $u$ is also a time series for each simulation for 
each component. If uncertainties are not known, this can be set to floating
point precision (which states that the metric is as precise as possible) or
some nominal fraction of the value (10\%, 20\%, etc). It is, of course, 
much preferred for the simulator to compute 
uncertainties directly. However, this is often not supported in the underlying
simulator. Choosing a nominal uncertainty, even if it is floating point 
precision, must suffice in such cases.

\begin{figure}[htb]
\centering
\includegraphics[width=0.45\textwidth]{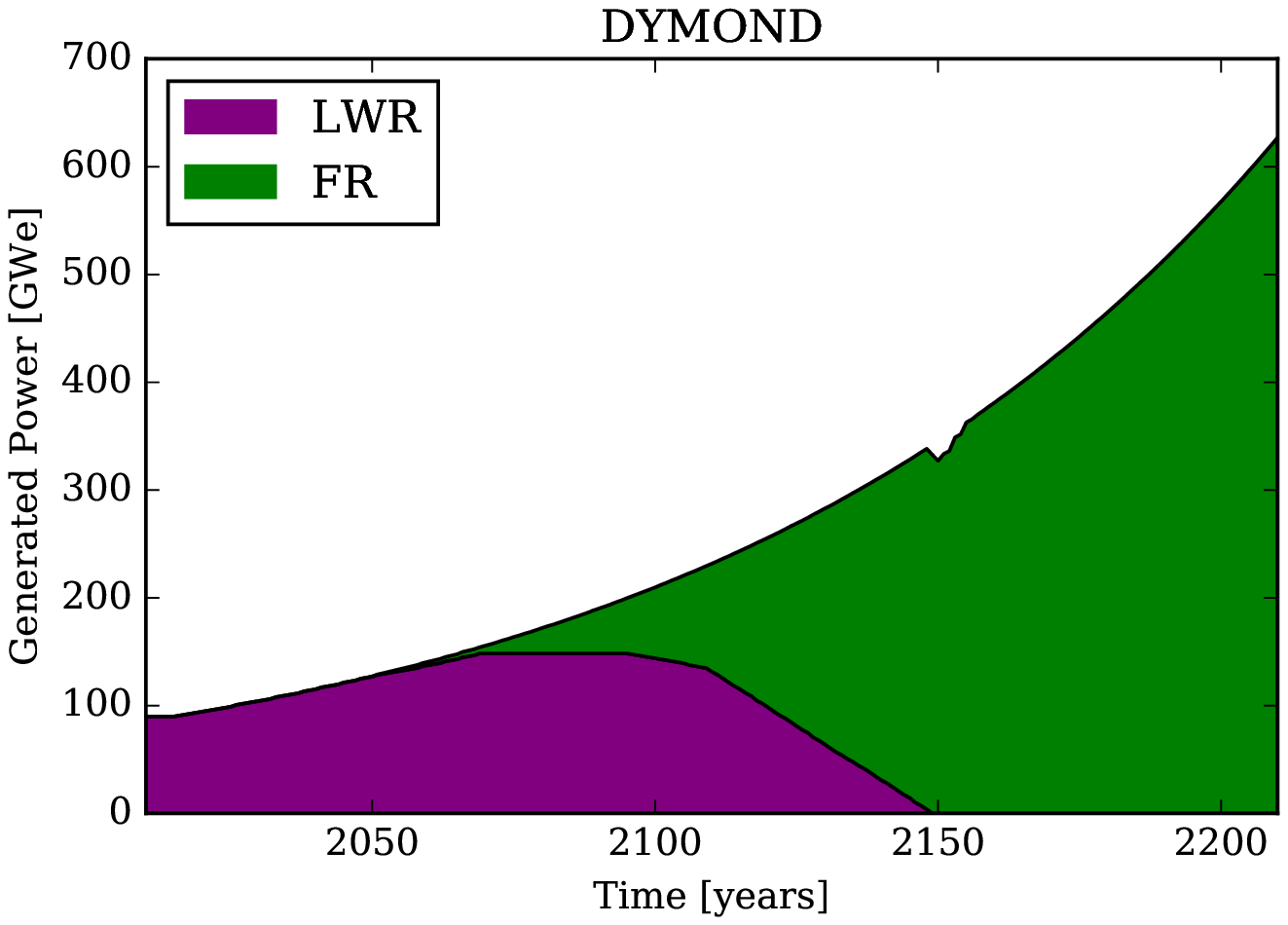}
\includegraphics[width=0.45\textwidth]{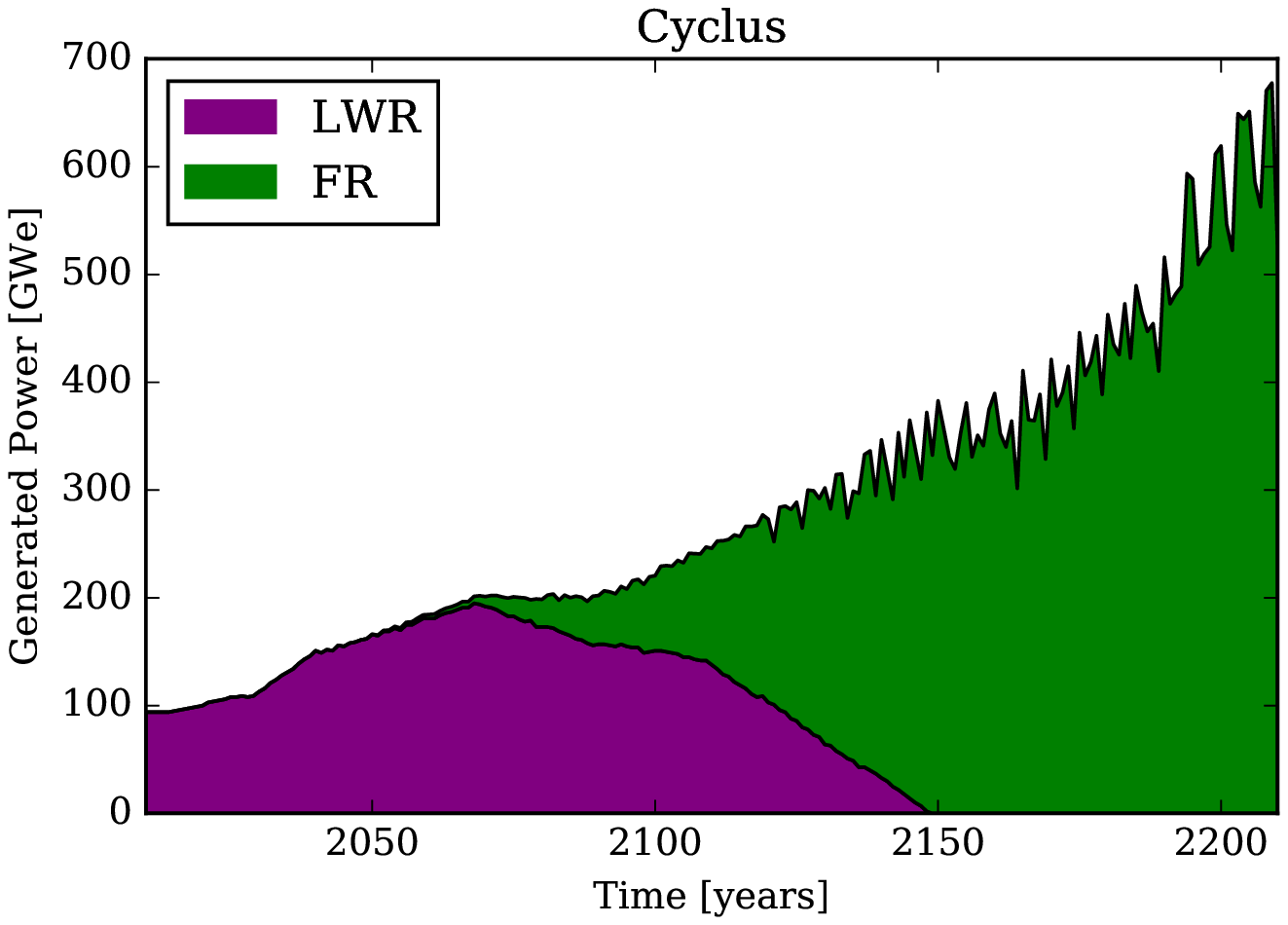}
\caption{The generated power in [GWe] as a function of time for the DYMOND and 
Cyclus simulators for both LWRs and FRs.}
\label{gwe-simulators}
\end{figure}

The demonstration here will use the generated power from LWRs and FRs in 
a transition that covers 200 years. The simulation should start with
90 GWe generated solely by LWRs and meet a 1\% growth in demand over the 
lifetime of the simulation. Figure \ref{gwe-simulators} shows the component time 
series curves for both DYMOND and Cyclus.  From here, the benchmarking 
analysis may proceed.

\clearpage

\section{Gaussian Process Modeling}
\label{gp}
The study of Gaussian processes is rich and deep and more completely covered by 
other resources, such as \cite{rasmussen2006gaussian}. Here only the barest of 
introductions to this topic are given as motivated by the regression problem at
hand. For nuclear fuel cycle benchmark analysis, Gaussian processes will be used
to form a model of the metric time series over all $S$ simulators. 

A Gaussian process is defined by mean and covariance functions. 
The mean function $\mu(t)$ is taken to be the expectation value $\E$ of 
the input functions, which here are the time series for all simulators. The
covariance function $k(t, t^\prime)$ is the expected value of the input 
functions to the mean. Symbolically, 
\begin{equation}
\label{mean-func}
\mu(t) = \E\left[m_s^i(t)\right] = \E\left[m_0^i(t), m_1^i(t), \ldots\right]
\end{equation}
\begin{equation}
\label{covar-func}
k(t, t^\prime) = \E\left[(m_s^i(t) - \mu(t))(m_s^i(t^\prime) - \mu(t^\prime))\right]
\end{equation}
The Gaussian process $\GP$ thus approximates the metric given information 
from all simulators. This is denoted either using functional or operator notation as follows:
\begin{equation}
\label{gp-def-approx}
m_*^i(t) \approx \GP\left(\mu(t), k(t, t^\prime)\right) \equiv \GP m_s^i
\end{equation}
Let $*$ indicate that the variable is associated with  
the model, rather than any of the raw simulator data.

For Gaussian process regression, a functional form for the covariance 
$k(t, t^\prime)$ needs to be provided.
This is sometimes called the kernel function and contains the free parameters 
for the regression, also called hyperparameters. How the hyperparameters are 
defined is 
tied to the definition of the kernel function itself. The values for the 
hyperparameters are determined
via optimization of the maximal likelihood of the value of the metric function.
While there are many possible kernels, a standard and generically applicable one 
is the exponential squared kernel, as seen in Equation \ref{exp2-kernel}:
\begin{equation}
\label{exp2-kernel}
k(t, t^\prime) = \sigma^2 \exp\left[-\frac{1}{2\ell}(t - t^\prime)^2 \right]
\end{equation}
Here, the length scale $\ell$ and the signal variance $\sigma^2$ are the 
hyperparameters.

Now, define a matrix $\K$ such that the element at the $t$-th row and $t^\prime$-th
column is given by Equation \ref{exp2-kernel}. If a vector of training metric 
values $\mathbf{m}$ is defined by concatenating metrics $m_s^i(t)$ for all 
simulators
for all times $T$, then log likelihood of the obtaining $\mathbf{m}$ 
is seen to be:
\begin{equation}
\label{log-p}
\log p(\mathbf{m}|T) = -\frac{1}{2}\mathbf{m}^\top\left(\K + u^2\I\right)^{-1}\mathbf{m}
                       -\frac{1}{2}\log\left|\K + u^2\I\right|
                       -\frac{n}{2}\log 2\pi
\end{equation}
Here, $u$ is the modeling uncertainty as denoted in the previous section, 
$\I$ is the identity matrix, and $n$ is the number of training points (the 
sum of the lengths of all of the time series). The hyperparameters $\ell$ and
$\sigma^2$ may be varied such that Equation \ref{log-p} is minimized. 
In this way, an optimal model for the metric is obtained.

Finally, suppose we want to evaluate the Gaussian process regression at a 
series of time points $\mathbf{t_*}$. 
The covariance vector between this time grid and the training data is denoted
as $\mathbf{k}_* = \mathbf{k}(\mathbf{t_*})$. The value of Gaussian process 
model of the metric function on the $\mathbf{t_*}$ time grid is thus seen to be:
\begin{equation}
\label{metric-model}
\mathbf{m}_*(\mathbf{t}_*) = \mathbf{k}_*^\top \left(\K + u^2\I\right)^{-1}\mathbf{m}
\end{equation}
A full derivation of Equations \ref{mean-func}-\ref{metric-model} can be found in 
\cite{rasmussen2006gaussian}. This resource also contains detailed discussions of 
how to optimize the hyperparameters, efficiently invert the covariance 
matrix, and compute the model values. 

In practice, though, a number of free and open source implementations of Gaussian 
process regression are readily available. In the Scientific Python ecosystem, both 
scikit-learn v0.17 \cite{scikit-learn} and George v0.2.1 \cite{hodlr} implement 
Gaussian process regression. For the remainder of this paper, George is used
for its superior performance characteristics and easier-to-use interface.

For purposes of nuclear fuel cycle benchmarking, Gaussian processes can be used to
create models of each component feature (e.g. LWRs and FRs) from the 
results of all simulators together. Other regression techniques could also be used 
to create similar models.  However, Gaussian processes have the advantage of 
incorporating modeling uncertainty, as seen above. The optimization of the 
hyperparameters yields the most accurate results for the model. Furthermore, the 
choice of kernel function can be tailored the functional form of the metric, if 
necessary. The exponential squared function was chosen because it is generic. If the 
metric is periodic, for instance, a cosine kernel may yield a more precise model. 
So while other regression techniques could be used, Gaussian processes 
encapsulate the correct behavior necessary for a non-judgmental benchmark while 
also remaining flexible to the particularities of the metric under examination.

\begin{figure}[htb]
\centering
\includegraphics[width=0.9\textwidth]{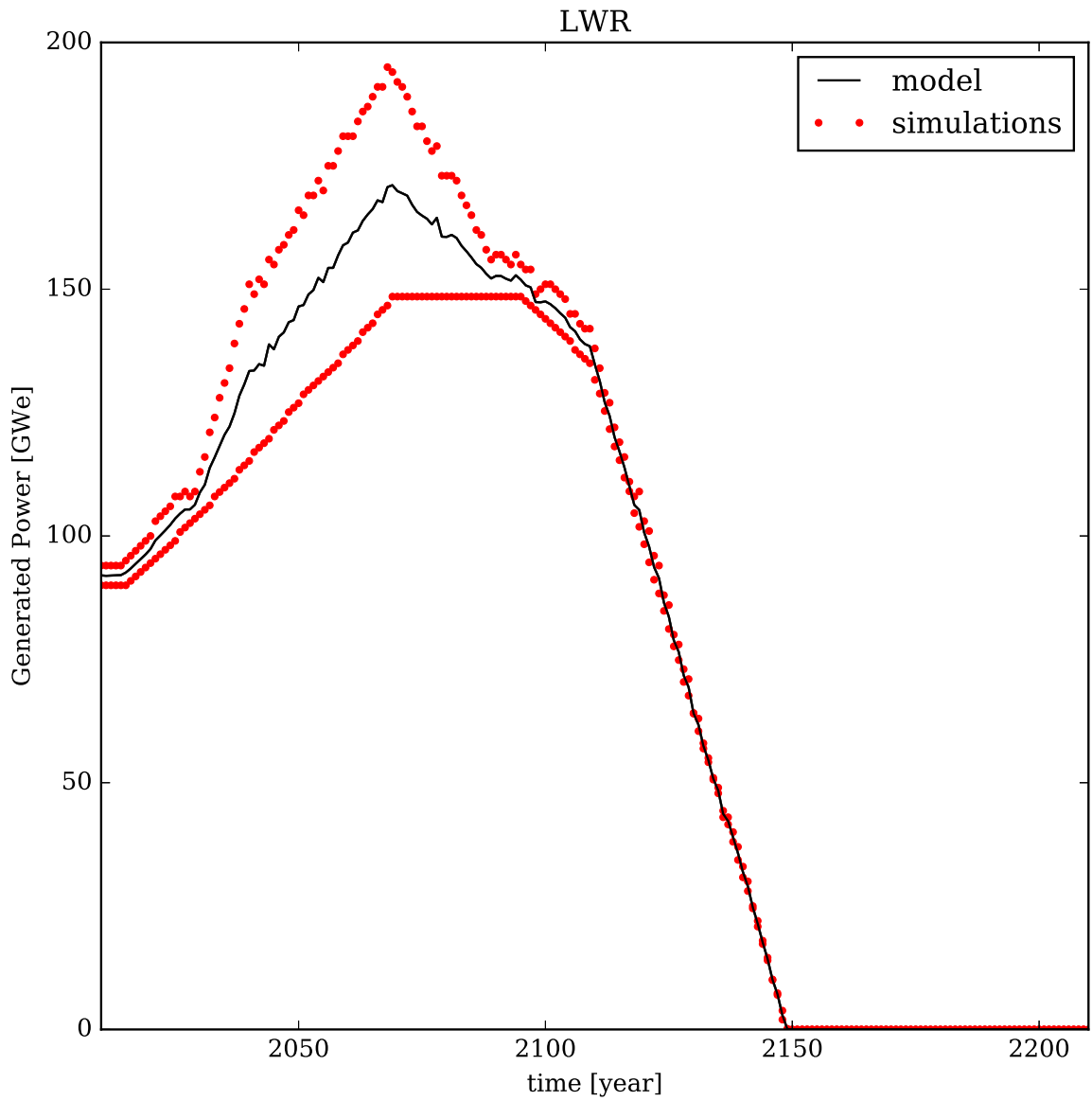}
\caption{The Gaussian process model of the generated power from LWRs [GWe] 
as a function of time as well as the results from the simulator that served as a 
training set for the model.}
\label{gwe-model-lwr}
\end{figure}

\begin{figure}[htb]
\centering
\includegraphics[width=0.9\textwidth]{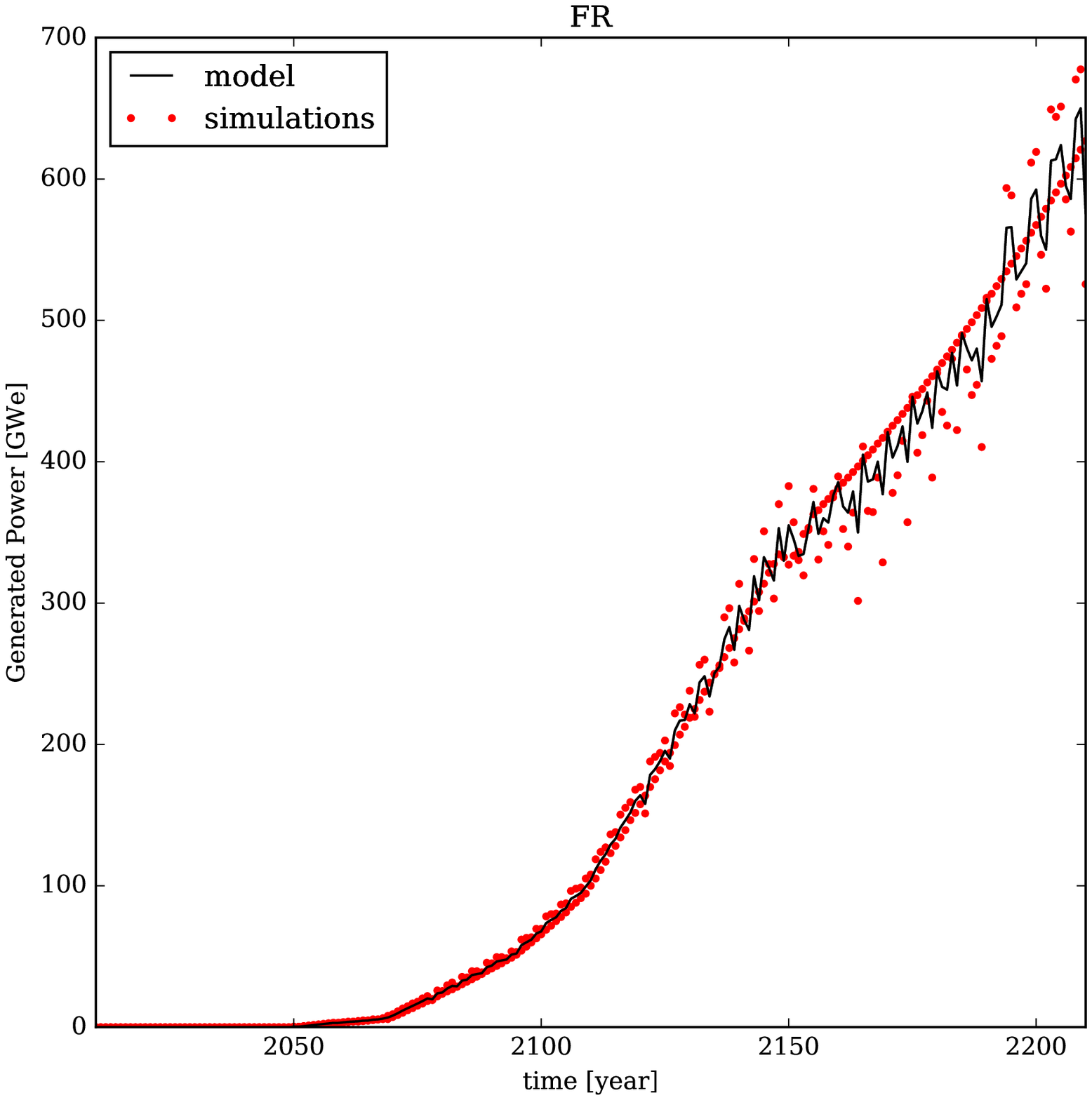}
\caption{The Gaussian process model of the generated power from FRs [GWe] 
as a function of time as well as the results from the simulator that served as a 
training set for the model.}
\label{gwe-model-fr}
\end{figure}

\begin{figure}[htb]
\centering
\includegraphics[width=0.9\textwidth]{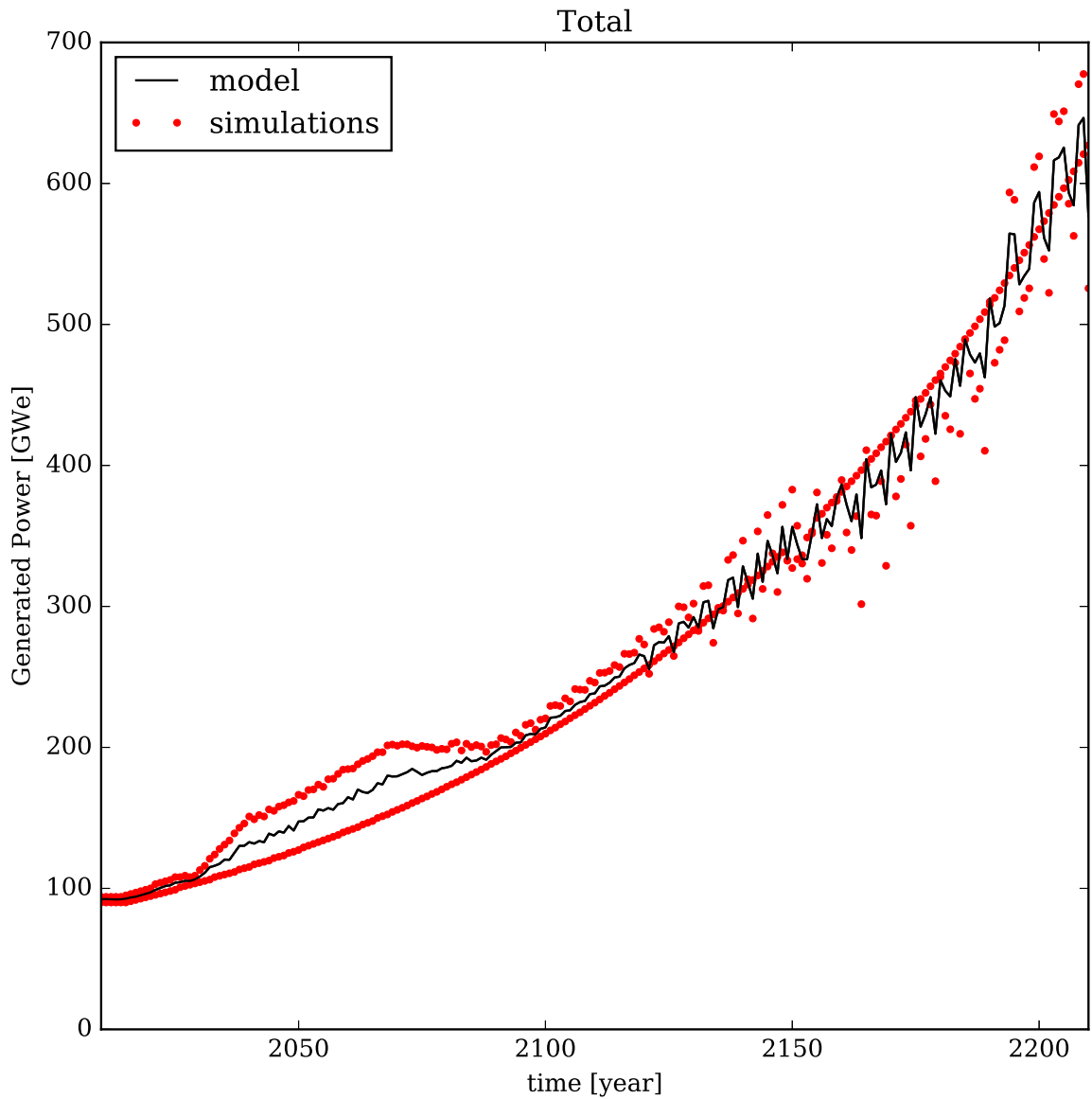}
\caption{The Gaussian process model of the total generated power [GWe] 
as a function of time as well as the results from the simulator that served as a 
training set for the model.}
\label{gwe-model-total}
\end{figure}

Using the sample data from \S\ref{setup}, three models can be constructed:
generated power from LWRs $m_*^{\LWR}(t)$, generated power from FRs $m_*^{\FR}(t)$, 
and total generated power $m_*(t)$. Assuming only floating point precision as 
the metric uncertainty, these Gaussian process models can be seen in Figures 
\ref{gwe-model-lwr} - \ref{gwe-model-total} along with the original time series 
from the simulators that train the model.
It is important to note that even though the sample data is only for two simulators, 
this method can handle any number of simulators, each with their own time grid, 
without modification.

\clearpage

Also note that Figure \ref{gwe-model-total} displays the model of the total 
generated power and not the total of the constituent models. Symbolically, 
\begin{equation}
\label{total-model}
m_* \approx \GP \left[\sum_i^I m_s^i\right] \ne \sum_i^I \GP m_s^i
\end{equation}
This is because the uncertainties are applied differently in these two cases. 
Moreover, the hyperparameter
optimization would not be consistent if one were to sum up constituent 
models. It is thus considered safer
to sum over the features for each simulator individually before applying the 
regression.

\begin{figure}[htb]
\centering
\includegraphics[width=0.9\textwidth]{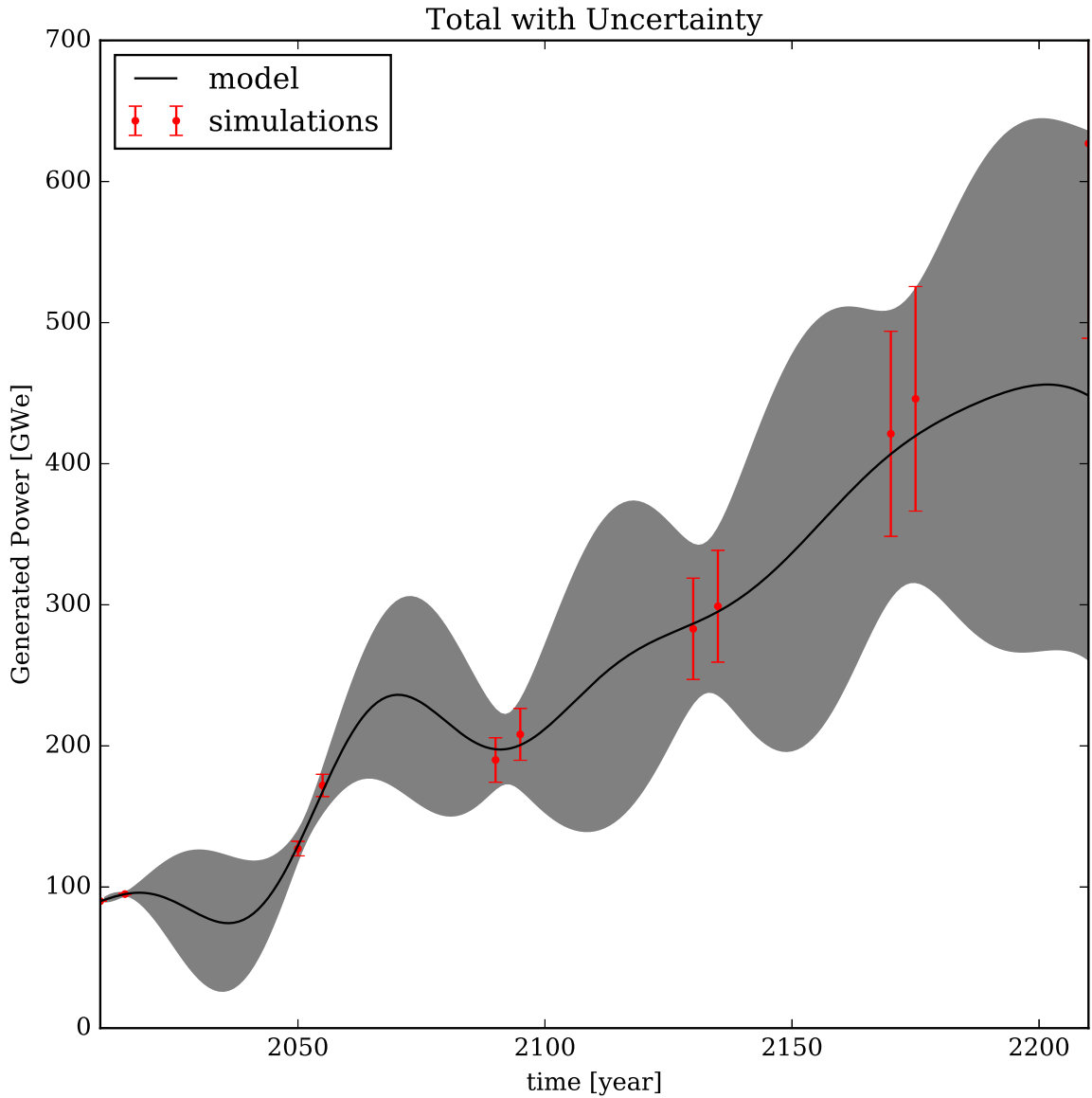}
\caption{The Gaussian process model of the total generated power [GWe] 
as a function of time. The training set data is relatively sparse and its uncertainty
has a 1\% per decade growth rate. The model curve is evaluated at every year. The 
gray envelope represents two standard deviations from the model, as computed by 
the Gaussian process.}
\label{gwe-model-total-with-uncertainty}
\end{figure}

Thus far the metric data has had effectively zero uncertainty.  But one of the 
desirable features of the Gaussian process regression is that it accounts for 
meaningful uncertainties. As a thought experiment, suppose that the time 
series data was much sparser and that 
the uncertainty associated with each value started off at zero and then grew at 
a rate of 1\% per decade. A model of the total generated power with this uncertainty 
is shown in Figure \ref{gwe-model-total-with-uncertainty}.

In Figure \ref{gwe-model-total-with-uncertainty}, note that as the model moves 
farther in time from the training data the standard deviation grows. Furthermore, 
as the uncertainty in the training data grows, the model itself degrades. Artifacts
from the choice of kernel begin to be visible in model whenever the uncertainties are 
relatively high. Both of these match intuition about how uncertain systems should
work.

Again, even though these uncertainty features are highly desirable in 
generic benchmarking applications, most nuclear fuel
cycle simulators do not report uncertainties along with the metrics they compute.
For this reason, the remainder of the examples in the paper will use the 
zero-uncertainty models as presented in Figures \ref{gwe-model-lwr} - 
\ref{gwe-model-total}.

\clearpage
\section{Dynamic Time Warping}
\label{dtw}

Now that there are representative models of all time series, the issue at hand is
how to compare these models. Dynamic time warping
is a method for computing the distance between any two time series. The time 
series need not be of the same length.  Furthermore, the time series may be 
decomposed into a set of representative spectra and the DTW may still be applied.
The distance computed by DTW is a measure of the changes that would need to be 
made to one time series to turn it into (warp) the other time series. Thus, 
the DTW distance is a measure over the whole time series, and not just a 
single characteristic point. As with Gaussian processes, a number of more thorough
resources, such as \cite{muller}, cover dynamic time warping in greater detail.
Here a minimal introduction is given that enables meaningful figure-of-merit 
calculations.

With respect to nuclear fuel cycle benchmarks, there are two main DTW applications.
The first is to compute the distance between a Gaussian process model and each of 
the simulators that made up the training set for that model. This gives a 
quantitative measure 
of how far each simulator is from the model and can help determine which 
simulators are outliers. To maintain a non-judgmental benchmark, though, it is 
critical to not then use this information to discard outliers.  Rather, outlier
identification should be used as part of an inter-code comparison. If one simulator
is an outlier for a given metric, the reasons for this should be investigated. 
For example, the outlier simulator may be at a higher fidelity level, there may be 
a bug in the outlier, or there may be a bug in all other simulators. Identifying 
outliers for many metrics could help discover the underlying cause of any 
discrepancies.

The second application of dynamic time warping to benchmarking is to compare 
the constituent 
feature models to the total model.  Distances computed in this manner allow 
for a rank ordering of the components.  This enables the benchmark to make claims
about which features drive the fuel cycle metric most strongly, over the whole 
simulation time domain and for all simulators. Traditionally, the simulators have 
to agree
within nominal error bounds ($<5\%$) for a benchmark to make such a claim.  Here,
the simulators need not necessarily agree since the Gaussian process models are
used as representatives.  In this application, it is useful to recast the DTW 
distance as a measure of contribution.  Contribution FOMs will be presented in 
\S\ref{contribution}.

For any two time series, dynamic time warping consists of three mathematical objects:
the distance $d$, a cost matrix $C$, and a warp path $w$. The cost matrix 
specifies how far a point on the first time series is from another point on the 
other time series.  The warp path is then the minimal cost curve through this 
matrix from the fist point in time to the last. The distance, therefore, is the
total cost of traversing the warp path.

The first step in a dynamic time warping algorithm is to compute the cost matrix. 
Suppose that the first 
time series $x$ has length $A$ indexed by $a$ and the second time series $y$ has 
length $B$ indexed by $b$. It is helpful to define an $A\times B$ matrix $\Delta L$
that is the $L_1$ norm of the difference of time series $x$ and $y$:
\begin{equation}
\label{delta-l1}
\Delta L_{a,b} = \left|x_a - y_b\right|_1
\end{equation}
The cost matrix $C$ is then an $A\times B$ sized matrix that is defined by the 
following recursion relations:
\begin{equation}
\label{cost-matrix}
\begin{split}
C_{1,1} & = \Delta L_{1,1}\\
C_{1,b+1} & = \Delta L_{1,b} + C_{1,b}\\
C_{a+1,1} & = \Delta L_{a,1} + C_{a,1}\\
C_{a+1,b+1} & = \Delta L_{a,b} + \min\left[C_{a,b}, C_{a+1,b}, C_{a,b+1}\right]
\end{split}
\end{equation}
The boundary conditions in Equation \ref{cost-matrix} are equivalent 
to applying an infinite cost to any $a$ or $b$ less than or equal to zero.
The units of the cost matrix are the same as the units of the metric. However, the 
scale of the cost matrix is geometrically larger than the metric itself. This is
due to the compounding nature of the recursive definition of $C$.

Given $C$, the warp path is thus a sequence of coordinate points that can then be 
computed by walking backwards through the matrix from $(A, B)$ to $(1, 1)$.
For a point $w_p$ in the warp path, the previous point $w_{p-1}$ is given by 
where the cost is minimized among the locations one column over $(a,b-1)$, 
one row over $(a-1,b)$, and one previous diagonal element to $(a-1,b-1)$. 
Symbolically, 
\begin{equation}
\label{warp-path}
w_{p-1} = \argmin\left[C_{a-1,b-1}, C_{a-1,b}, C_{a,b-1}\right]
\end{equation}
The maximum possible length of $w$ is thus $A + B$ and the minimum length is 
$\sqrt{A^2 + B^2}$. The warp path itself could potentially serve as a FOM.  
However, doing so would not take into account the cost along this path.

Therefore, the distance $d$ is defined as a FOM which does include for the cost of the
warp.  Due to the recursion relations used to define the cost matrix, $d$ can be 
stated succinctly as:
\begin{equation}
\label{d-calc}
d = \frac{C_{A,B}}{A + B}
\end{equation}
That is, $d$ is the final value of the cost matrix divided by the maximal length 
of the warp path.

For all practical purposes in nuclear fuel cycle benchmarking, $A$ and $B$ can be 
forced to have the same
value. This is because the Gaussian process model can be used to predict a time series 
with whatever time grid is desired.  The advantage of using a regression model 
is that it allows the analyst to force the same time grid.  The advantage of 
dynamic time warping is that ensuring the same time grid is not necessary.
Coupling Gaussian processes and DTW together is a more robust analysis tool 
than the methods individually.

\begin{figure}[htb]
\centering
\includegraphics[width=0.9\textwidth]{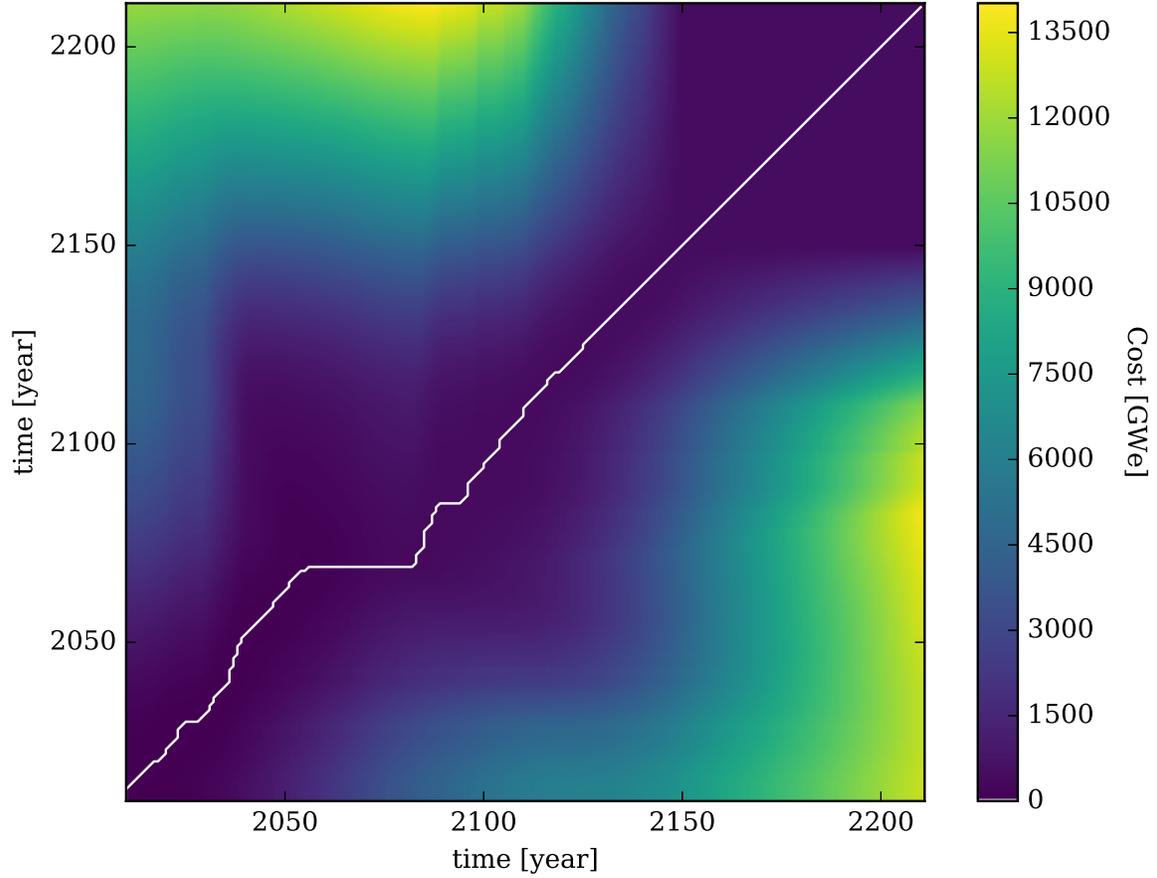}
\caption{Heat map of the cost matrix between the Gaussian process model 
for LWRs, $m_*^\LWR(t)$, and the Cyclus LWR time series, $m_\CYCLUS^\LWR(t)$.
The warp path is superimposed as the white curve on top of the cost matrix.}
\label{cost-lwr-model-to-lwr-cyclus}
\end{figure}

Figure \ref{cost-lwr-model-to-lwr-cyclus} displays an example cost matrix 
as a heat map for the DTW between the LWR Gaussian process model 
$m_*^\LWR(t)$ and the original Cyclus LWR time series $m_\CYCLUS^\LWR(t)$.
Additionally, the warp path between these two is shown as the white curve
on top of the heat map. Note that while $w$ is monotonic along both time axes, the
path it takes minimizes the cost matrix at every step. Higher cost regions
have the effect of pushing the warp path along one axis or another. The 
distance between these two curves $d(m_*^\LWR, m_\CYCLUS^\LWR)$ is computed 
to be 1.053 GWe.

The process of dynamic time warping model generated data to raw simulator data can be 
repeated for all combinations of simulators and features. The 
$d(m_*^i, m_s^i)$ that are computed may then directly serve as a FOM for
outlier identification. Only distances between the same feature $i$ may be compared,
as they share the same model. For instance, it is valid to compare
$d(m_*^\LWR, m_\DYMOND^\LWR)$ and $d(m_*^\LWR, m_\CYCLUS^\LWR)$. However,  
it is not valid to compare $d(m_*^\LWR, m_\DYMOND^\LWR)$ and 
$d(m_*^\FR, m_\DYMOND^\FR)$.  In the situation, where the simulators 
have different time grids, the model can be evaluated with each time grid
prior to computing the corresponding DTW distances and the comparison will
remain valid.

\begin{table}[htb]
\centering
\caption{Distances [GWe] between models and simulators for all combinations of 
simulators (DYMOND and Cyclus) and metric features (generated power for 
LWRs, FRs, and in total). The simulators are presented in the rows and the
features are given as columns. Distances may only be compared along 
each column.}
\label{d-compare}
\begin{tabular}{l||c||c||c|}
                & \textbf{LWR} & \textbf{FR} & \textbf{Total} \\
\hline
\textbf{DYMOND} & 1.452        & 2.783       & 3.022          \\
\hline
\textbf{Cyclus} & 1.053        & 3.732       & 3.984          \\
\hline
\end{tabular}
\end{table}

Table \ref{d-compare} gives the distances for all simulator and feature 
combinations in the sample data presented in \S\ref{setup}. Only the differences between DYMOND and 
Cyclus for the same feature may be directly compared.  However, Table \ref{d-compare} 
does indicate that the DYMOND is closer to the model since both the FR and Total
generated power distances are closer for DYMOND than for Cyclus.  As many
simulators as desired could be added to the benchmark and distances could 
be tallied for them as well. At a sufficient number of simulators, usual 
statistics (mean, standard deviation) along each column may be computed.
Simulators whose metrics fall outside of a typical threshold of the mean
(one or two standard deviations) would then be considered outliers and 
subject to further inter-code comparison. The two simulators here are 
for demonstration purposes and neither can be said to be an outlier in a
non-judgmental way. Recall, though, that outliers determined in this way 
may stem from the more correct simulator. The term outlier is not a 
condemnation on its own.

The second application of DTW to nuclear fuel cycle benchmarking is as a
FOM for contribution of constituent features to a total metric. For this application, 
compute the distance between the model of the total and the model of the 
part, namely $d(m_*^\Total, m_*^i)$ for all $i \in I$. This provides a 
measure of how much the total metric is determined by a particular part
over the whole time series for all simulators.
Using this measure is only reasonable if the total metric is known 
to be the sum of its constituents.  This assumption is valid for 
a large percentage of nominal benchmarking metrics. In particular, 
mass flows and generated power both follow this rule. Metrics that are 
linear transformations of these basic metrics (such as decay heat or 
radiotoxicity) will also conform to this constraint. The distances computed here
can then be used to rank order the importance of each constituent feature. 
Smaller distances are closer to the total and thus more important.

\begin{figure}[htb]
\centering
\includegraphics[width=0.9\textwidth]{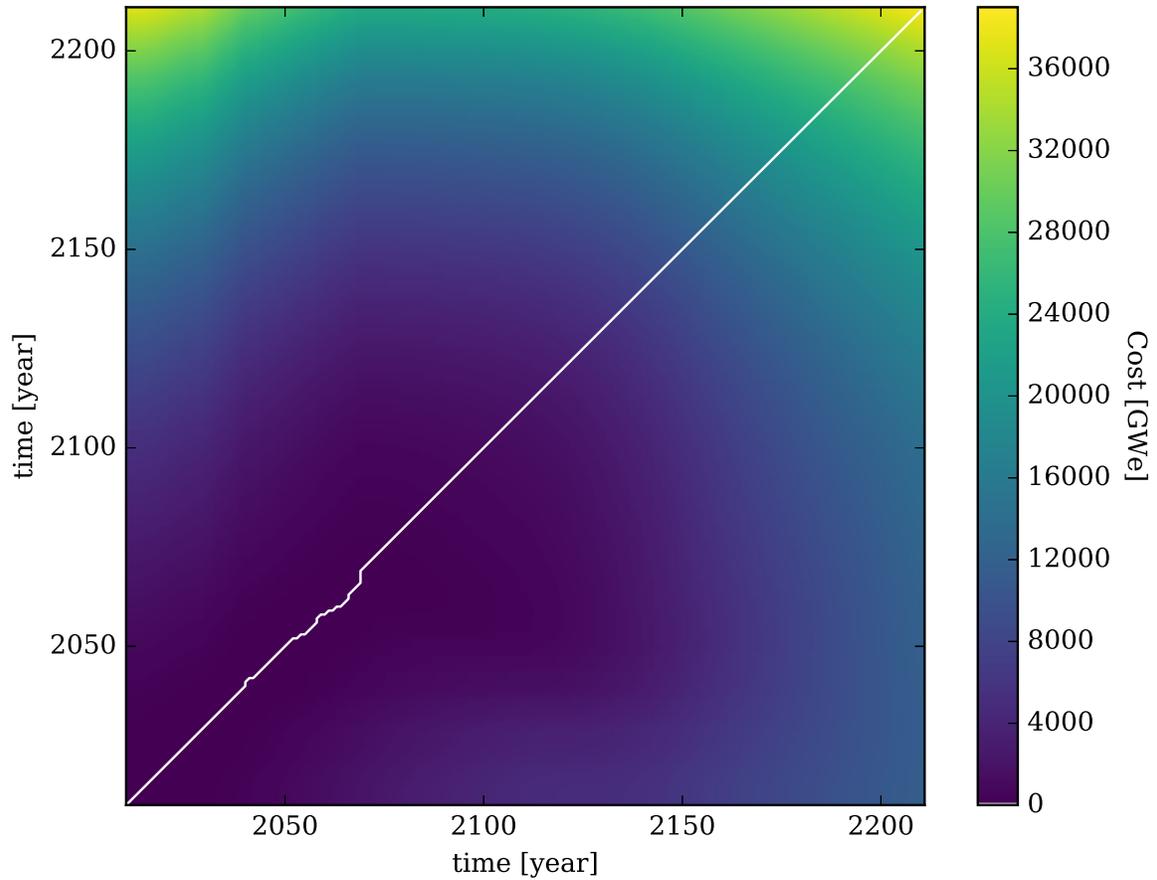}
\caption{Heat map of the cost matrix between the Gaussian process model 
for total generated power, $m_*^\Total(t)$, and the LWR model, 
$m_*^\LWR(t)$.
The warp path is superimposed as the white curve on top of the cost matrix.}
\label{cost-total-model-to-lwr-model}
\end{figure}

\begin{figure}[htb]
\centering
\includegraphics[width=0.9\textwidth]{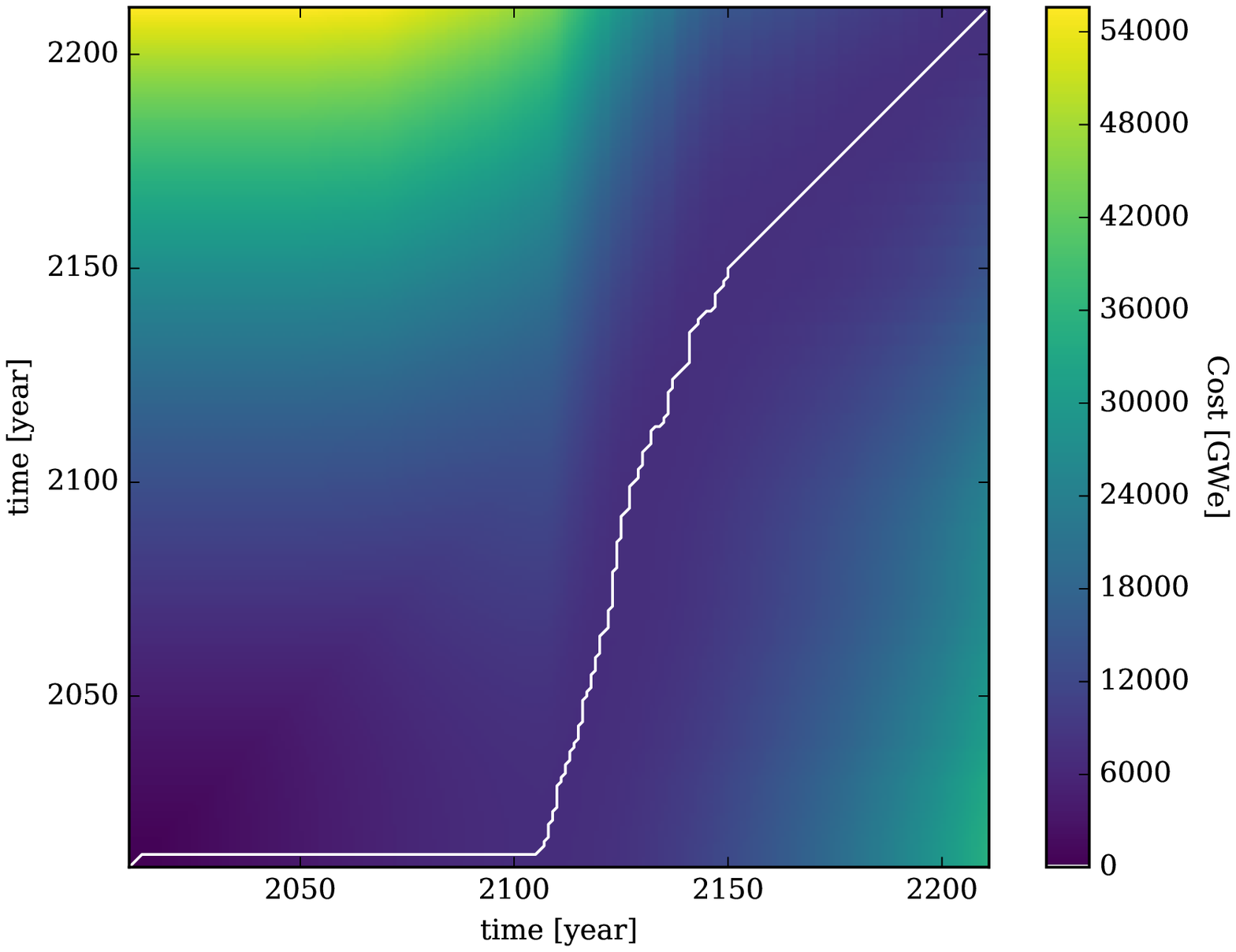}
\caption{Heat map of the cost matrix between the Gaussian process model 
for total generated power, $m_*^\Total(t)$, and the FR model, 
$m_*^\FR(t)$.
The warp path is superimposed as the white curve on top of the cost matrix.}
\label{cost-total-model-to-fr-model}
\end{figure}

In the sample data here, the total generated power model can be compared to 
the models for LWR and FR generated power. The value of 
$d(m_*^\Total, m_*^\LWR) = 97.010$ while $d(m_*^\Total, m_*^\FR) = 19.503$.
As expected, the FRs are a more important driver of the transition system 
than the LWRs over the time domain examined. Figures 
\ref{cost-total-model-to-lwr-model} \& \ref{cost-total-model-to-fr-model}
show the cost matrices and warp paths for these two cases.  Notice that in the
FR case, the warp path is effectively flat until the FRs are deployed in 
significant numbers. 
 
In summary, dynamic time warping yields a meaningful mechanism for benchmarks to 
compare various simulators and models. As a tool for
outlier identification, it allows for each simulator to use its own native 
time grid. As a tool for performing rank ordering of features, the DTW distances
are a perfectly functional FOM. However, lower distances implying higher 
importance may run counter to intuition. The potential for confusion, therefore,
makes it less than ideal as a FOM on its own. A remedy for this is presented
in the next section in form of a new contribution FOM.

\clearpage
\section{Contribution}
\label{contribution}

In \S\ref{dtw}, the dynamic time warping distance was presented as a 
FOM for measuring the similarity between time series, whether they 
were model-to-simulator comparisons or model-to-model comparisons.
The distances alone, though, have the inverse meaning when trying to 
compute a FOM for contribution to a fuel cycle.  For example, the 
question may be posed as, ``Are LWRs or FRs more important to the fuel 
cycle as a whole?'' While distances can answer this question, smaller 
disantances have higher impact. Preferably, a higher FOM score would yield a 
higher impact. Furthermore, the distances in the previous section have
the same units as the cost matrix and are similarly bounded. A better 
FOM for contribution would be unitless and defined on the range $[0,1]$.

Thus, let $c$ be the \emph{contribution} FOM that satisfies the above 
constraints. To define $c$, first define $D$ as the maximal possible 
distance from the model of the total. Recall that this time series has length $A$.
$D$ is, therefore, the L1 norm of the model of the total time series divided
by twice its length. 
\begin{equation}
\label{D-def}
D(m_*^\Total) = \frac{\left|m_*^\Total\right|_1}{2A}
\end{equation}
This is the equivalent to computing the DTW distance
between $m_*^\Total$ and the curve where the metric is zero for all time 
(i.e. the t-axis itself).  This relies on the notion that 
metric is necessarily non-negative everywhere.  If the metric is allowed to 
be negative, another baseline curve could be chosen. $D$ would then be 
computed as the DTW distance between the total model and this baseline.
However, in most cases the metrics are not allowed to be negative, 
a baseline of zero is suitable, and Equation \ref{D-def} applies.

The contribution figure-of-merit of a given partition to the total is thus 
defined as follows:
\begin{equation}
\label{cont}
c^i = 1 - \frac{d(m_*^\Total, m_*^i)}{D(m_*^\Total)}
\end{equation}
$D$ is seen to normalize the model-to-model distance while subtracting this
ratio from unity makes higher contribution values more important.  
Using the sample data, $c^\LWR = 0.298$ and $c^\FR = 0.859$. This again shows
that the FRs are more important to the total power of the whole cycle.
Here though, higher contribution scores yield higher importance and the values
are always between zero and one.

Note that even though $c$ is a fraction, it is not normalized across 
partitions. Namely, the sum of all contributions for all $I$ partitions
is on the following range, which is not $[0, 1]$:
\begin{equation}
\label{sum-c-range}
0 \le \sum_i^I c^i \le I
\end{equation}
It is easy to imagine an alternative FOM that divides $c^i$ by $I$. However, 
this was not done here because the choice of $I$ (the number of partitions)
can be made arbitrarily large.  In the sample calculations $I=2$ for LWRs and
FRs.  However, $I$ could have been set to 3 by including small modular reactor
(SMRs) which are never built and produce no power.  $I$ could then be 
increased to 4 or higher by including more non-existent reactor types.
Dividing by $I$ is not stable enough for a FOM.

Furthermore, dividing the contribution by $I$ is not sufficient to 
normalize the sum of the $c^i$, in general.  This is because the 
contribution is inherently a cumulative measure. If a component ever had a 
non-zero value, it will always be seen to have contributed something. 
Because of this constraint on the total, $\sum c^i$ can never
reach $I$ unless there is only a single component or the 
total is zero valued everywhere (which implies that the constituents are also 
zero). Thus, dividing $c^i$ by $I$ with the aim of normalizing the 
sum of the $c^i$ is incorrect for any situation of interest to a fuel 
cycle benchmark.

If a truly normalized version of contribution is desired, it must 
use the sum of the actual contributions. Define the normalized contribution 
as $|c^i|$, 
\begin{equation}
\label{norm-ci}
\left|c^i\right| = \frac{c^i}{\sum_j^I c^j}
\end{equation}
The disadvantage with the normalized contribution is that all of the 
individual component $c^i$ must be known prior to the normalization. 
Furthermore, since the sum is typically greater than 1, the difference 
between components is often smaller in the normalized form than in the
more pronounced unnormalized contribution.
The only advantages that $|c^i|$ confers over $c^i$ are that it is defined on 
the range $[0,1]$ and that $\sum |c^i| = 1$. Otherwise, both versions of the FOM
have the same properties.

\begin{figure}[htb]
\centering
\includegraphics[width=0.9\textwidth]{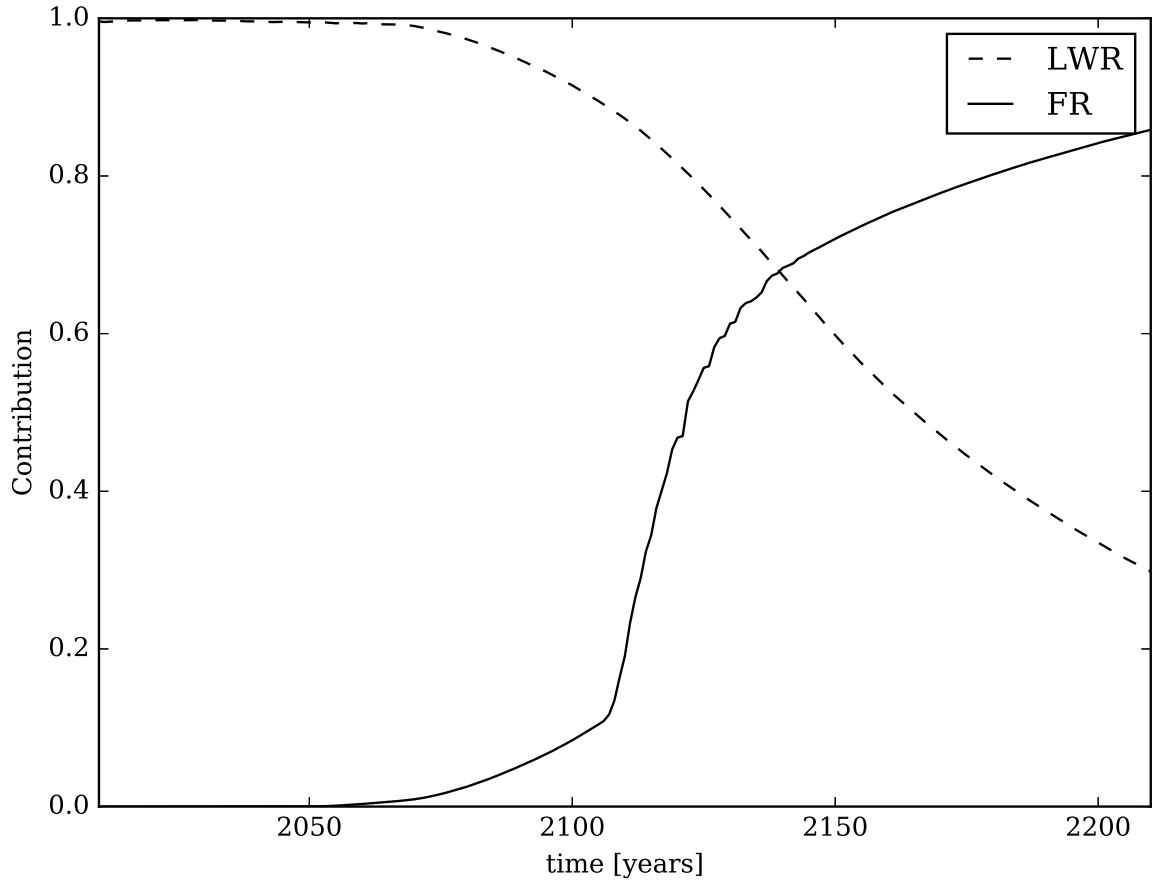}
\caption{Contribution of LWRs and FRs to total generated power as a 
function of time.}
\label{c-of-t}
\end{figure}

\begin{figure}[htb]
\centering
\includegraphics[width=0.9\textwidth]{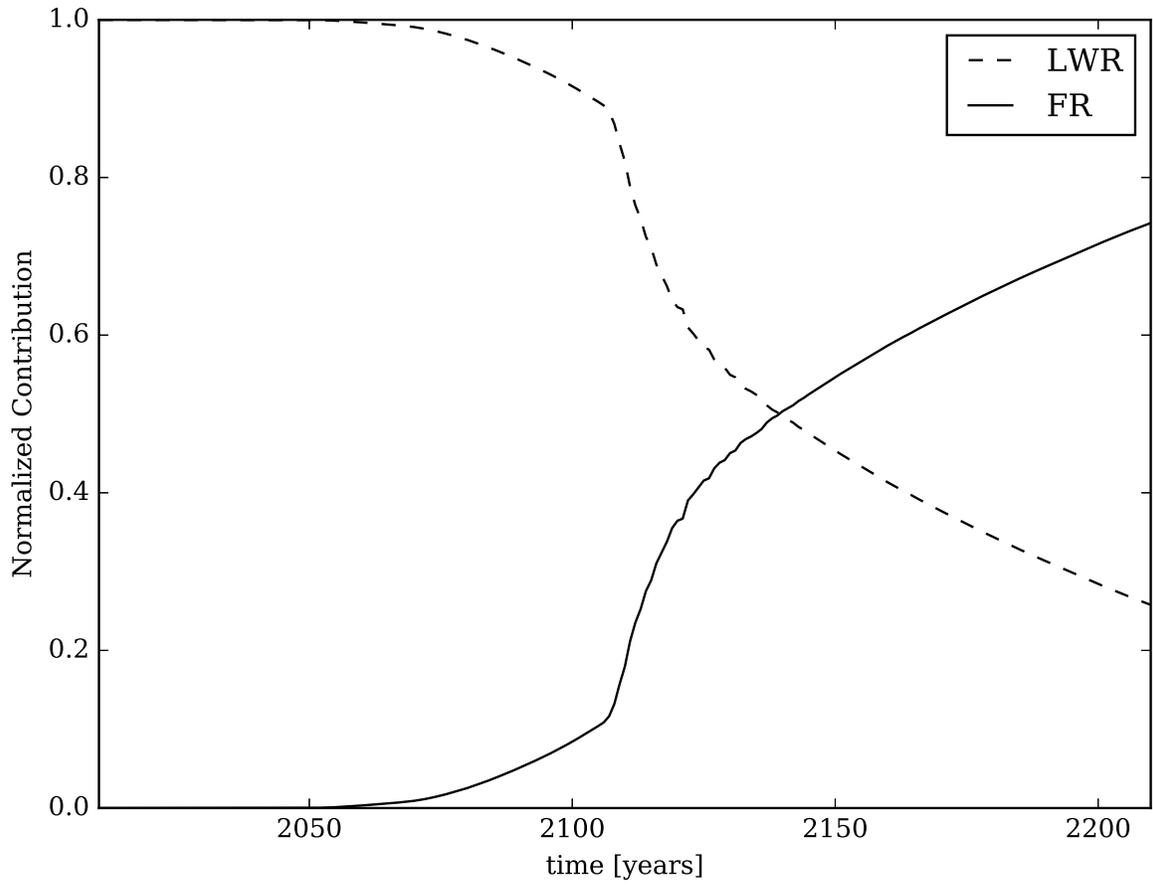}
\caption{Normalized contribution of LWRs and FRs to total generated power 
as a function of time.}
\label{normc-of-t}
\end{figure}

Both $c^i$ and $|c^i|$ can be viewed as a function of time.
Doing so in a nuclear fuel cycle benchmark may help identify artifacts in the
calculation that are the result of the time domain chosen for the benchmark itself.
Figures \ref{c-of-t} \& \ref{normc-of-t} display the contribution and 
normalized contribution respectively for both LWRs and FRs over the full 
time domain. From these figures, the point where FRs become more important 
than LWRs is seen to be approximately year 2140.

\clearpage
\section{Cautionary Tale on Filtering}
\label{filtering}

It is tempting to insert standard filtering techniques from signal processing 
after creating a Gaussian process model but prior to any dynamic time warping 
calculations. A fast Fourier transform (FFT) based low-pass filter 
\cite{merletti1999standards,moreland2003fft} or 
Mel-frequency cepstral coefficients (MFCC) \cite{muda2010voice,imai1983cepstral} could be used to reduce error 
in the model itself, 
and thus make the contribution FOM more precise. Unfortunately, most 
fuel cycle metrics are not well-formed candidates for such filtering strategies.
Including such filters as part of the analysis can easily lead to wildly unphysical
models.

Consider a simple low-pass filter where a 256 channel real-valued FFT frequency 
transform is taken.  All but lowest 32 channels are discarded prior to the applying 
the inverse transform. High frequency jitter in the original signal is removed, 
allowing for a better signal-to-noise ratio.

\begin{figure}[htb]
\centering
\includegraphics[width=0.9\textwidth]{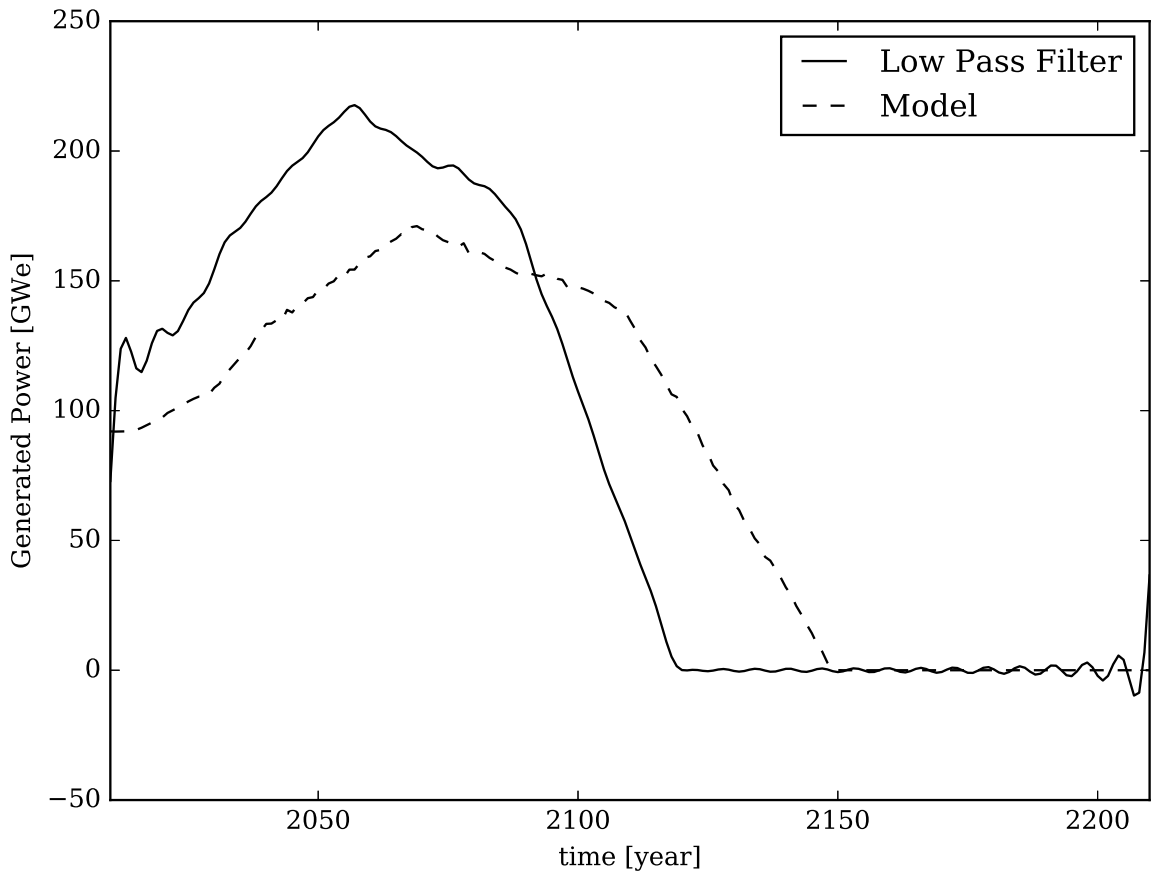}
\caption{Low-pass FFT filter of LWR Gaussian process model $m_*^\LWR$ alongside
the unfiltered model itself.}
\label{fft-lwr-model}
\end{figure}

Figure \ref{fft-lwr-model} shows the results of applying the low-pass filter 
described above to the Gaussian process model of the LWR generated power, 
$m_*^\LWR(t)$.  The filtered curve demonstrates at least three major problems.  The
first is that the values of the curve are allowed to be negative, which is 
impossible for this (and many other) fuel cycle metrics.  The second is that 
near the time boundaries ($t=2010$ and $t=2210$), the amplitude of the filtered model
is significantly higher than the unfiltered model. At $t=2210$, the metric should be zero but
instead is 36.5 GWe. Thirdly, the shape of the curve itself is skewed to lower 
times. The time at which the metric goes to zero should be near year 2150 but is 
instead closer to year 2115.  All of these issues would severely distort any 
DTW calculations that follow.

The reason behind these inconsistencies is that the FFT process is fundamentally 
periodic.  However, using the annual time grid here, the LWR generated power metric
is not periodic. Neither is the modeling error for most fuel cycle metrics periodic
on an annual basis. 
Thus, while well-intentioned, a low-pass filter is not generally applicable.

Alternatively, MFCCs provide a mechanism for converting a time series into a 
set of power spectrum coefficient curves. Since the dynamic time warping procedure
uses an L1 norm to form the cost matrix, the MFCCs of two signals can be directly 
compared. Each coefficient should roughly correspond in shape and amplitude to some
feature in the original signal.  Noisy, high frequency coefficients tend to be 
very similar and so their contribution to a DTW distance is correspondingly less 
than the contribution for lower mode coefficients. Coupling MFCC to DTW is an 
extremely common method employed in speech recognition systems 
\cite{muda2010voice,milner2002speech,sato2007emotion}.  

\begin{figure}[htb]
\centering
\includegraphics[width=0.9\textwidth]{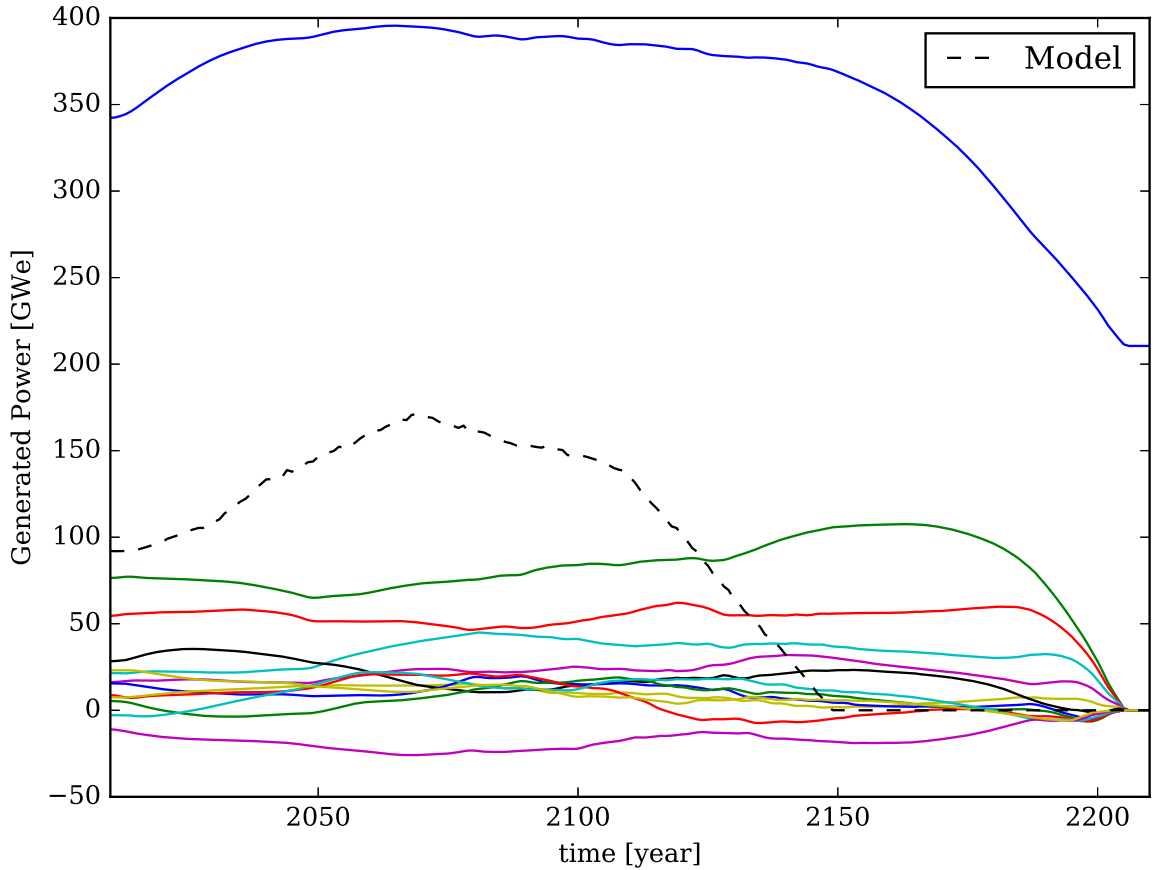}
\caption{Representative Mel-frequency cepstral coefficients (solid lines) of an 
LWR Gaussian process model $m_*^\LWR$ alongside the model itself (dashed line).}
\label{mfcc-lwr-model}
\end{figure}

Figure \ref{mfcc-lwr-model} displays the MFCC curves of the LWR Gaussian 
process model as well as the model itself. None of these curves, not even the major 
coefficient, resembles the actual model.  Rather, the cosine basis for MFCCs
is clearly visible.  As with the low-pass filter, the MFCCs also have 
uncharacteristic negative components.  Moreover, 
the metric data is not sampled frequently enough to have meaningful
time windows. For the fuel cycle metrics here, there is only one data point per year 
and the signal itself may change in a meaningful way each year. By comparison, 
in speech recognition, audio is sampled at least at 22050 Hz for characteristic 
signals on the order of 1 second.  The data volume for fuel cycle benchmark metrics
is simply too low for MFCC transformations to capture the desired features. 

Proceeding anyways, suppose the contribution measure is computed for the MFCCs 
of LWR, FR, and 
total generated power models.  In this case the LWR contribution is found to be 
0.572 while the FR contribution is 0.899. Using the models directly as was done 
in \S\ref{contribution}, the contribution values were 0.298 and 0.859 respectively.
This implies that using the MFCCs had the opposite effect as desired.  The MFCCs
added error to the FOM and made the LWR and FR contributions seem more alike than 
they truly are.

Therefore filtering the models prior to dynamic time warping is a dubious practice
in the general case. In all likelihood, the metric does not meet the underlying 
assumptions of the filter. The metric may not be periodic or may not be sampled 
frequently enough. Sometimes it may be possible to construct a metric that does
meet these qualifications. For instance, the generated power could be sampled monthly  
such that seasonal demand behavior is noticeable. Even in such a case, it is instead recommended
to pick a different kernel for the Gaussian process model such that these 
periodic behaviors are captured.  The regression itself then takes on the role of 
minimizing model uncertainty. Further filtering to this end becomes redundant and
dangerous.  Additionally, it is unlikely that 
the majority of the simulators would be able to calculate such a high-fidelity metric.
That alone should disqualify such metrics or FOMs from any benchmarking study or
inter-code comparison.

\section{Conclusions \& Future Work}
\label{conclusion}

This paper demonstrates a robust method for generating figures-of-merit
for nuclear fuel cycle benchmarking activities by coupling Gaussian process
regression to dynamic time warping. This method takes advantage of modeling
uncertainties in fuel cycle metrics if they are known. It is also capable 
of handling the situation where different simulators output metric data on
vastly different time grids. The distance computed by the dynamic time 
warping can itself serve as the figure-of-merit. Additionally, the 
distance can also be used to derive contribution and normalized contribution
figures-of-merit.

Any regression method could have been used to form a model. Similarly, any
mechanism for comparing two time series could have been used as a measure
of distance.  However, Gaussian processes and DTW were chosen because of 
the nature of a benchmarks and inter-code comparisons that lack experimental
validation. It is not possible to build out a given fuel cycle scenario
and see how it performs 200 years in the future. Furthermore, using 
historical data for validation provides too few cases for comparison and 
each simulator could simply be tuned to precisely match historical events.
Thus, each simulator in a benchmark could be valid or they all could be 
invalid. It is therefore necessary for the FOM to not skew for or against 
any particular simulator. Gaussian process models as used here do not 
judge the simulators differently. The DTW then takes into account the 
cumulative effect of the whole time domain and does not preferentially 
select certain times.

The sample benchmark presented here was very simple and was used for motivation 
purposes only. It consisted of just
two simulators (DYMOND and Cyclus) and one metric (generated power) with
two components (LWR and FR).  However, both Gaussian processes and DTW
are inherently multivariate. More complex forms of analysis could therefore
be performed. For example, the Gaussian process could jointly model the 
effect from many inputs onto the metric. Perhaps the benchmark is formulated
to look at the generated power as a function of time and the power demand curve.
In this case, a two dimensional GP model would be used. Alternatively, 
suppose that a matrix time series of the all individual nuclide mass flows 
are available. DTW is still able compute the distance between two 
such matrices. This would yield a measure of how the mass flows themselves
differ - taking into account each nuclide component - without rely on a collapsed
one dimensional total mass flow curve.  Such cases will be considered in
future work as real inter-code comparison data becomes available.

Furthermore, this work focused on the particular use case of benchmarking.
However, the FOM calculations presented here could also be used to evaluate 
different fuel cycle scenarios. DTW distances could be computed between
a business-as-usual once through scenario and an LWR-to-FR transition
scenario, or any other proposed scenario. This provides a measure for 
comparing the relative cost (in units of the metric, not necessarily 
economic) for selecting one cycle over another. The work here, thus, 
should be seen as a stepping stone to further fuel cycle scenario evaluation
work.

Lastly, dynamic time warping could itself serve a purpose as the objective 
function in a fuel cycle optimization.  For example, suppose a power demand curve 
such as 1\% power growth is known. The DTW distance from the total generated
power to this curve could be minimized as a function of the reactor 
deployment schedule. Such a distance could potentially yield a more 
precise or faster optimization process than simply taking the sum of 
the differences between two time series. Such an optimization would also allow
for matching on multiple time series features simultaneously while retaining
a real-valued objective function.

\section*{Acknowledgements}
\label{acknow}
The author would like to express deep gratitude to Dr. Bo Feng of
Argonne National Lab for providing the DYMOND data used throughout this 
paper.

\bibliographystyle{ans}
\bibliography{refs}
\end{document}